\documentclass[12pt]{article}

\usepackage[
top    = 2.50cm,
bottom = 2.50cm,
left   = 2.50cm,
right  = 2.50cm]{geometry}
\usepackage{jheppub}
\usepackage[sort&compress]{natbib}

\usepackage{amssymb}
\usepackage{subcaption}
\usepackage{graphicx}
\usepackage{slashed}
\usepackage{amsmath}
\usepackage{hyperref}
\usepackage{verbatim}
\usepackage{caption}
\usepackage{subcaption}
\usepackage{graphicx}

\newcommand{\be}{\begin{equation}}
\newcommand{\ee}{\end{equation}}
\newcommand{\bea}{\begin{eqnarray}}
\newcommand{\eea}{\end{eqnarray}}
\newcommand{\beq}{\begin{equation}}
\newcommand{\eeq}{\end{equation}}
\newcommand{\beqa}{\begin{eqnarray}}
\newcommand{\eeqa}{\end{eqnarray}}
\newcommand{\beqar}{\begin{eqnarray*}}
\newcommand{\eeqar}{\end{eqnarray*}}

\begin{document}

\title{Gravitating Scalars and Critical Collapse in the Large $D$ Limit}

\author{Moshe Rozali,}
\author{Benson Way}

\affiliation{
\it{
Department of Physics and Astronomy, University of British Columbia,\\
Vancouver, BC V6T 1Z1, Canada}}

\emailAdd{rozali@phas.ubc.ca}
\emailAdd{benson@phas.ubc.ca}

\abstract{We develop the large $D$ limit of general relativity for spherically symmetric scalar fields in both asymptotically flat and  asymptotically anti-de Sitter spaces.  The leading order equations in the $1/D$ expansion can be solved analytically, providing a large $D$ description of oscillating soliton stars. When the amplitude reaches a critical threshold, certain divergences occur which we interpret as signal of horizon formation. We estimate the size of the resulting black hole and obtain, with respect to that definition, a Choptuik scaling exponent for our family of solutions.
}

\maketitle

\vspace{10pt}
\section{Introduction}

The large dimension limit of General Relativity \cite{Emparan:2013moa} offers substantial simplification in various circumstances.  Difficult numerical calculations at finite $D$ can be reproduced at large $D$ with significantly fewer computational resources, or even sometimes analytically.  Examples include the instability of rotating black holes \cite{Emparan:2014jca, Andrade:2018nsz}, the Gregory-Laflamme instability \cite{Emparan:2015gva, Rozali:2016yhw, Dandekar:2016jrp, Emparan:2018bmi}, holographic turbulence \cite{Rozali:2017bll}, and more.

Currently, the large $D$ limit of general relativity has mostly been developed for black holes.  It was observed that the physics of black holes and black branes simplifies in an appropriate scaling limit in which the number of transverse directions is taken to be large \cite{Emparan:2014cia, Emparan:2014aba, Emparan:2015rva}. In such a limit, many interesting physical questions localize to the horizon of the black hole, where they can be investigated using an effective field theory on the black hole ``membrane" (see \cite{Emparan:2015hwa, Emparan:2016sjk} and \cite{Bhattacharyya:2015dva, Bhattacharyya:2015fdk,Dandekar:2016fvw,Bhattacharyya:2016nhn,Dandekar:2017aiv}).  Effectively, the large $D$ expansion acts as a gradient expansion, in which variations in directions perpendicular to the horizon are consistently sub-leading in the expansion. It is thus quite similar to the fluid/gravity correspondence.

It is interesting then to attempt to utilize a large dimension expansion (perhaps with different scaling limits) to provide similar simplification to gravitational phenomena that are not localized to a pre-existing black hole horizon. Examples of possible such contexts include super-radiant instabilities, holographic superconductivity \cite{Emparan:2013oza}, driven turbulence, black hole mergers, and many mother interesting contexts that could perhaps benefit from the underlying insight of \cite{Emparan:2013moa}.

In this paper, we develop the large $D$ limit of general relativity in situations where horizons may or may not exist.  Such situations include strongly gravitating objects like oscillaton stars and boson stars, as well as phenomena like critical collapse and the instability of anti-de Sitter space (AdS). We will focus on the simplest and most well-studied models exhibiting these phenomena, which consists of a gravitating scalar field with spherical symmetry, either in asymptotically flat space or asymptotically AdS spaces.

Let us briefly review what is known for these models at finite $D$, beginning with critical collapse \cite{Choptuik:1992jv}.  Consider a real, massless scalar field with spherical symmetry that interacts gravitationally in asymptotically flat space.  Dynamically, the scalar field tends to either disperse or collapse into a black hole.  One can fine-tune the initial data so as to be near the critical point of gravitational collapse.  Near criticality, evolution approaches that of a universal critical solution before eventually dispersing or collapsing.  This critical solution exhibits discrete self similarity. That is, there is a coordinate $\tau$ such that the metric satisfies $g=e^{-2\tau}\tilde g$, where $\tilde g$ is a metric that is periodic in $\tau$ with period $\Delta$.  For collapsing data near criticality, the mass of the black hole that is created scales as
\begin{equation}
M\propto (p-p^*)^\gamma\;,
\end{equation}
where $p$ parametrises the initial data with $p^*$ being the critical value, and $\gamma$ is a constant.   Both the echoing period $\Delta$ and the critical exponent $\gamma$ are universal in the sense that they are the same for all one-parameter families of data that pass through criticality.  These constants can be obtained numerically, either by either time evolution \cite{Choptuik:1992jv} (done up to $D=14$ \cite{Garfinkle:1999zy,Sorkin:2005vz,Bland:2005vu}) or by directly constructing the critical solution and perturbing it (done in $D=4$ \cite{Gundlach:1996eg}). 

Besides critical phenomena, there are also quasi-stationary solutions in asymptotically flat space. For a massive real scalar, there are oscillating soliton stars, sometimes called oscillatons or oscillons \cite{Seidel:1991zh}, where the scalar field and metric oscillate periodically. For a complex scalar field, there are similar configurations called boson stars \cite{Lee:1991ax,Schunck:2003kk,Liebling:2012fv}, where only the phase of the scalar field oscillates.  These are one-parameter families of solutions which can be parametrised by the value of the scalar field at the origin $\varphi_0$.  There is a critical value of $\varphi_0>\varphi_0^*$ where oscillatons or boson stars become unstable. $\varphi_0$ does not seem to have a bound, and it appears that the limit $\varphi\to\infty$ is singular, where the scalar and metric curvature diverge. 

Unlike Minkowski space, (global) AdS contains a reflecting boundary that allows for an arbitrarily small excitations to form black holes \cite{DafermosHolzegel2006,Dafermos2006,Bizon:2011gg}.  This is the celebrated AdS instability. However, there is also a large class of initial data that do not appear to form black holes \cite{Dias:2012tq,Maliborski:2013jca,Buchel:2013uba,Balasubramanian:2014cja,Bizon:2014bya,Balasubramanian:2015uua,Dimitrakopoulos:2015pwa,Green:2015dsa}.  The separation between these two types of initial data remains poorly understood, although there is growing evidence \cite{Dimitrakopoulos:2015pwa,Choptuik:2018ptp} that the non-collapsing data is intimately connected to oscillatory solutions, which are the non-linear extensions of the normal modes of AdS, such as oscillons \cite{Maliborski:2013jca,Fodor:2015eia}, boson stars \cite{Dias:2011at,Liebling:2012fv,Buchel:2013uba,Choptuik:2017cyd}, and geons \cite{Dias:2011ss,Horowitz:2014hja,Martinon:2017uyo}.

The outline of the paper is as follows. Using our large $D$ expansion we make connection to this discussion, for both the asymptotically flat and asymptotically AdS case, which we separately discuss in the two following sections. In both cases we construct a family of horizonless strongly gravitating scalar field ``stars", which we call oscillatons. These solution exist in asymptotically $AdS$ space for finite $D$, but for the asymptotically flat case their existence is surprising, and may not extend (as absolutely stable  objects) to finite values of $D$. For the asymptotically $AdS$ case we can also discuss the extension to certain multi-mode solutions, which we predict to be long lived in the large $D$ limit. 

Our family of solution is characterized by an amplitude, and when it reaches a critical value, certain divergences occur which we take as a signal for horizon formation. When the amplitude is close to that threshold, we can estimate the size of the resulting black hole to obtain a Choptuik scaling exponent. It is unclear though how this exponent is related to the finite $D$ Choptuik scaling exponent, defined in the conventional way. However, the numerical value we find sees to be the expected critical exponent  at large $D$, as we discuss below.

After discussing the asymptotically flat case in section 2, and the asymptotically $AdS$ case in section 3, we conclude with a summary and possible directions for future research.

\vspace{10pt}
\section{Large $D$ Scalars in Flat space}
We now develop the large $D$ limit of a scalar field in the asymptotically flat case, which is technically simpler than the asymptotically $AdS$ case, to be discussed in the next section. 
\vspace{10pt}
\subsection{Probe Scalar in Minkowski Spacetime}
To gain intuition on the nature of the large $D$ limit we are taking, we start by considering a probe real scalar field in flat spacetime, under spherical symmetry:
\begin{equation}\label{eq:flatprobe}
\partial_t^2\varphi=\partial_r^2\varphi+\frac{D-2}{r}\partial_r\varphi-m^2\varphi\;.
\end{equation}

We know the set of solutions to this equation for all $D$, but let us attempt to find an approximate solution via a large $D$ expansion.  The equation \eqref{eq:flatprobe} suggests that in order to have nontrivial equations in the large $D$ limit we need to perform a rescaling of time to $\tau=\sqrt{D-2} t$. Intuitively, the large $D$ limit causes spheres of slightly different radii to be very different from each other, resulting in high frequency oscillations, and so must be compensated by rescaling of time. 

Performing this rescaling and a large $D$ expansion gives at lowest order
\begin{equation}\label{eq:largedflatprobe}
\partial_\tau^2\varphi_0=\frac{1}{r}\partial_r\varphi_0\;.
\end{equation} 
Note that the resulting equation is parabolic, with the roles of space and time swapped. This will be the case for all equations we obtain in the large $D$ limit, below.  Note also that the mass $m$ does not appear in these equations, since we have opted not to scale it with $D$. The large $D$ scalar equation at lowest order is effectively massless, and we henceforth only consider massless scalar fields.

\begin{figure}
\centering
\includegraphics[width=.7\textwidth]{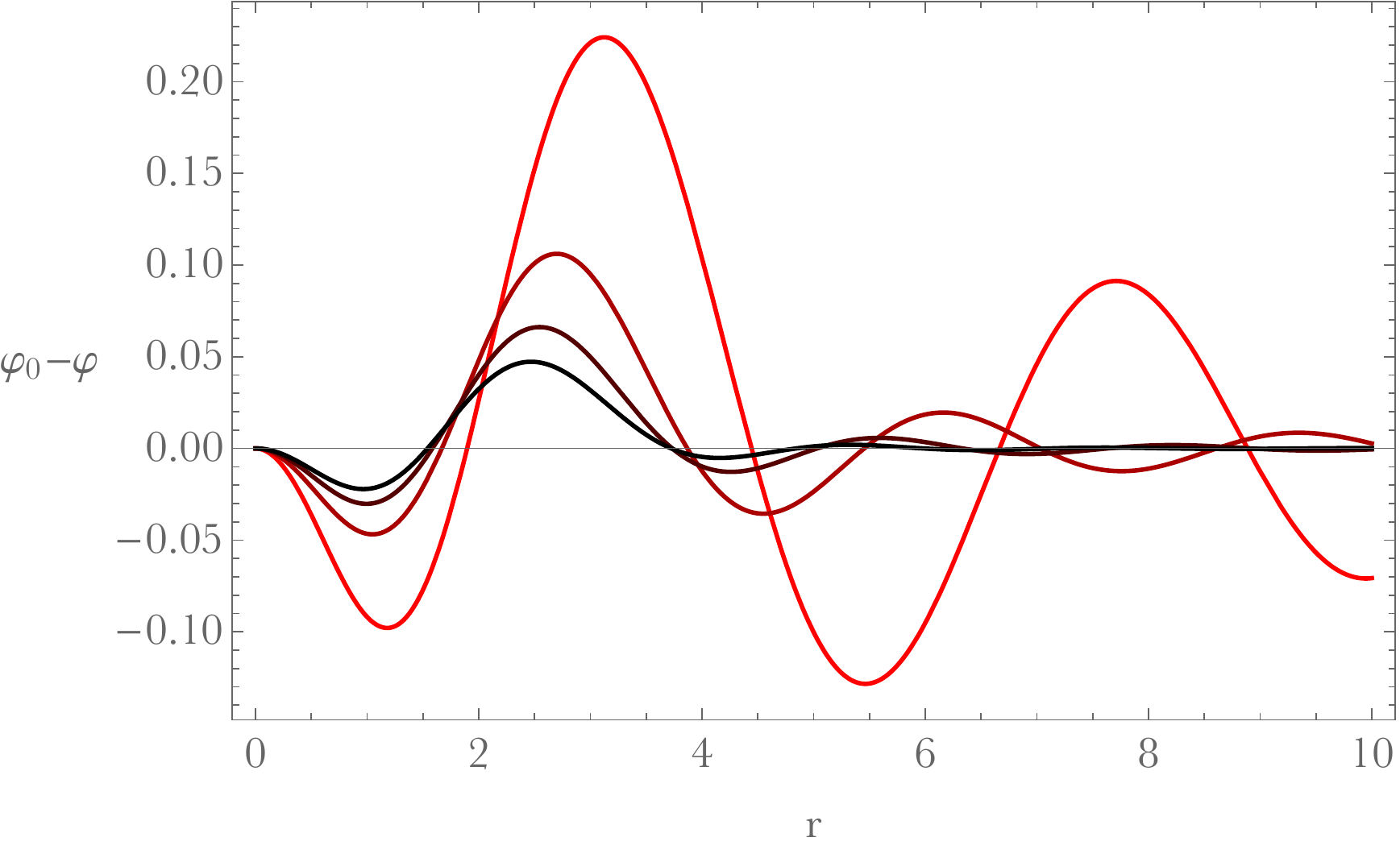}
\caption{Comparison of the exact solution $\varphi$ to the large $D$ solution $\varphi_0$ for $D=4,6, 8, 10$. We see that the approximation improves with increasing $D$. }
\label{fig:bessel}
\end{figure}

 The solutions to \eqref{eq:largedflatprobe}, obtained via separation of variables, can be written as
\begin{equation}\label{eq:largedflatprobesol}
\varphi_0=a_\omega\cos(\omega \tau+\phi_0)e^{-\frac{1}{2}\omega^2r^2}\;.
\end{equation}
We can then compare the result to the exact solutions of \eqref{eq:flatprobe} with $m=0$ which are given by (a sum of) the Bessel functions
\begin{equation}
\varphi=a_\omega\cos(\omega \tau+\phi_0)r^{-\frac{D-3}{2}}J_{\frac{D-3}{2}}(\sqrt {D-2}\omega r)\;.
\end{equation}
Scale invariance and time translation invariance lets us, without loss of generality, consider $\omega=1$, and $\phi_0=0$. We normalise the functions so that $\varphi(0,r=0)=1$ at the origin. The difference $\varphi_0-\varphi$ at $\tau=0$ is then shown in Fig. \ref{fig:bessel}.
It is reassuring that the approximation improves with increasing $D$.  

\vspace{10pt}
\subsection{Large D scalars in Asymptotically Flat Space}
Now let us apply the same approximation to a scalar field with backreaction.  We choose a Schwarzschild gauge where the metric can be written
\begin{equation}
\mathrm ds^2=-(1-\delta)A\,\mathrm dt^2+\frac{\mathrm dr^2}{A}+r^2 \mathrm d\Omega_{D-2}\;.
\end{equation} 
Where all functions depend on the radial coordinate $r$ and time $t$. The full equations of motion can be written
\begin{subequations}
\begin{align}
\partial_t A&=-\frac{2r}{D-2}A\partial_t\varphi\partial_r\varphi\;,\\
\partial_r A&=\frac{D-3}{r}(1-A)-\frac{r}{D-2}\left[\left(\frac{\partial_t\varphi}{\sqrt{1-\delta}A}\right)^2+(\partial_r\varphi)^2\right]A\;,\\
\partial_r\delta&=-\frac{2r}{D-2}\left[\left(\frac{\partial_t\varphi}{\sqrt{1-\delta}A}\right)^2+(\partial_r\varphi)^2\right](1-\delta)\;,\\
0&=-\frac{1}{\sqrt{1-\delta}A}\partial_t\left(\frac{\partial_t\varphi}{\sqrt{1-\delta}A}\right)+\frac{1}{r}\partial_r(r\partial_r\varphi)+\frac{D-3}{r A}\partial_r\varphi\;.
\end{align}
\end{subequations}

Now we move to the time coordinate $\tau=\sqrt D t$ and perform an expansion in $1/D$:
\begin{equation}\label{flatexpansion}
A=1-\frac{1}{D}A_0+O(D^{-2})\;,\qquad \delta=\delta_0+O(D^{-1})\;,\qquad \varphi=\varphi_0+O(D^{-1})\;.
\end{equation}
where the lowest order term in $A$ is fixed by the equations of motion. Note that in this expansion, $A_0 \ll D$ implies that within our approximation $A\neq0$. Therefore, strictly speaking, horizon formation does not occur at infinite $D$.  However, as we shall see, there are solutions where $A_0$ diverges.  We conjecture that the divergence in those solutions is a signal of horizon formation at large but finite $D$. 

An expansion of the equations of motion then gives, to leading order
\begin{subequations}\label{eq:largedflat}
\begin{align}
A_0&=r^2\frac{(\partial_t\varphi_0)^2}{1-\delta_0}\;,\\
\partial_r \delta_0&=-2r(\partial_\tau\varphi_0)^2\;,\\
\partial_r\varphi_0&=\frac{r}{\sqrt{1-\delta_0}}\partial_\tau\left(\frac{\partial_\tau\varphi_0}{\sqrt{1-\delta_0}}\right)\;.
\end{align}
\end{subequations}
Note that the metric function $A_0$ has decoupled, reducing the equations to a nonlinear system in $\delta_0$ and $\varphi_0$.

Like the probe scalar equation at large $D$ \eqref{eq:largedflatprobe}, the scalar field equation resembles a heat equation with spatial and temporal coordinates swapped.  This raises the question of how such an equation should be solved.  Analogy with the heat equation suggests that `initial data' $\varphi_0(\tau,r=0)$ should be given, and then integrated to larger values of $r$.  Without loss of generality, one can choose the condition for the metric $\delta_0(\tau,r=0)=0$. After obtaining a solution, one can use invariance of the equations under $\delta_0\rightarrow\delta_0+c(\tau)$ to shift $\delta_0$ to have the more standard condition $\delta_0(\tau,r\to\infty)=0$.  The nature of the heat equation and the linear probe solution \eqref{eq:largedflatprobesol} suggest that $\varphi_0$ will decay exponentially at large $r$. 

If the scalar field and its derivatives remains finite under this construction, so too will the metric functions. This will generate a `non-collapsing' solution to the field equations at leading order in $1/D$.  Otherwise, if the scalar field or its derivatives diverge, then $A_0$ will also diverge, leading to a breakdown of the large $D$ equations, possibly indicating horizon formation at finite $D$.

One can continue the expansion by including higher order terms in \eqref{flatexpansion}, and expanding the equations of motion further. One will then obtain linear equations in the higher-order variables that are sourced nonlinearly by the lower order solutions. 

\vspace{10pt}
\subsection{Oscillatons and Choptuik Scaling}
Fortunately, and surprisingly, there is an exact analytic solution to the equations \eqref{eq:largedflat}, given by
\begin{equation}
\varphi_0=\tanh^{-1}(\epsilon \,  e^{-\omega^2r^2/2}\sin(\omega\tau-\tau_0))\;,\qquad \delta_0=\frac{\epsilon^2 \, e^{-\omega^2r^2}\cos^2(\omega\tau-\tau_0)}{1-\epsilon^2  \, e^{-\omega^2r^2}\sin^2(\omega\tau-\tau_0)}\;,
\end{equation}
for some amplitude $\epsilon$, phase $\tau_0$, and frequency $\omega$.  Using that solution we also have
\begin{equation}
A_0=\frac{\epsilon^2 \,\omega^2 r^2e^{-\omega^2r^2}\cos^2(\omega\tau-\tau_0)}{(1-\epsilon^2 \, e^{-\omega^2r^2})(1-\epsilon^2 \, e^{-\omega^2r^2}\sin^2(\omega\tau-\tau_0))}\;.
\end{equation}
One can use scale invariance to fix $\omega$, and translation invariance to fix $\tau_0$, leaving a one-parameter family of solutions parametrised by the amplitude $\epsilon$. 

This family of solutions has a sharp threshold at  $\epsilon = 1$. For $\epsilon<1$ the solution is regular and the scalar field oscillates indefinitely.  Since this oscillation is a single frequency, this solution describes the large $D$ version of an oscillaton.  However, we note that at finite $D$, oscillatons in asymptotically flat space only exist for massive scalar fields.   This large $D$ solution might have an alternative interpretation for finite $D$ massless fields as states that are especially long-lived. 

For $\epsilon\geq1$, the scalar field diverges. For any $\epsilon>1$, $A_0$ also diverges. As discussed above, we conjecture that this divergence is a signal of horizon formation at finite $D$.  Let us use this divergence to estimate a critical exponent.  In the usual treatment of Choptuik scaling, we fine tune to the threshold for black hole formation, and look for the mass of the resulting small black hole. Note however that the exponential falloff of $A_0$ implies that the our large $D$ solutions have zero energy, we therefore cannot use a mass as a measure for the critical exponent.   Instead, to get a measure for the size of the black hole above the threshold for collapse, we look the radius where $A_0$ diverges, i.e. the location of the putative horizon. The divergence occurs when $\epsilon^2 \, e^{-r^2}=1$.  An expansion about $\epsilon=1$ yields
\begin{equation}
r_\mathrm{div}=\sqrt{2(\epsilon-1)}+O((\epsilon-1)^{3/2})\;,
\end{equation}
implying a critical exponent $\gamma=1/2$.  

\begin{figure}[h]
\centering
\includegraphics[width=.6\textwidth]{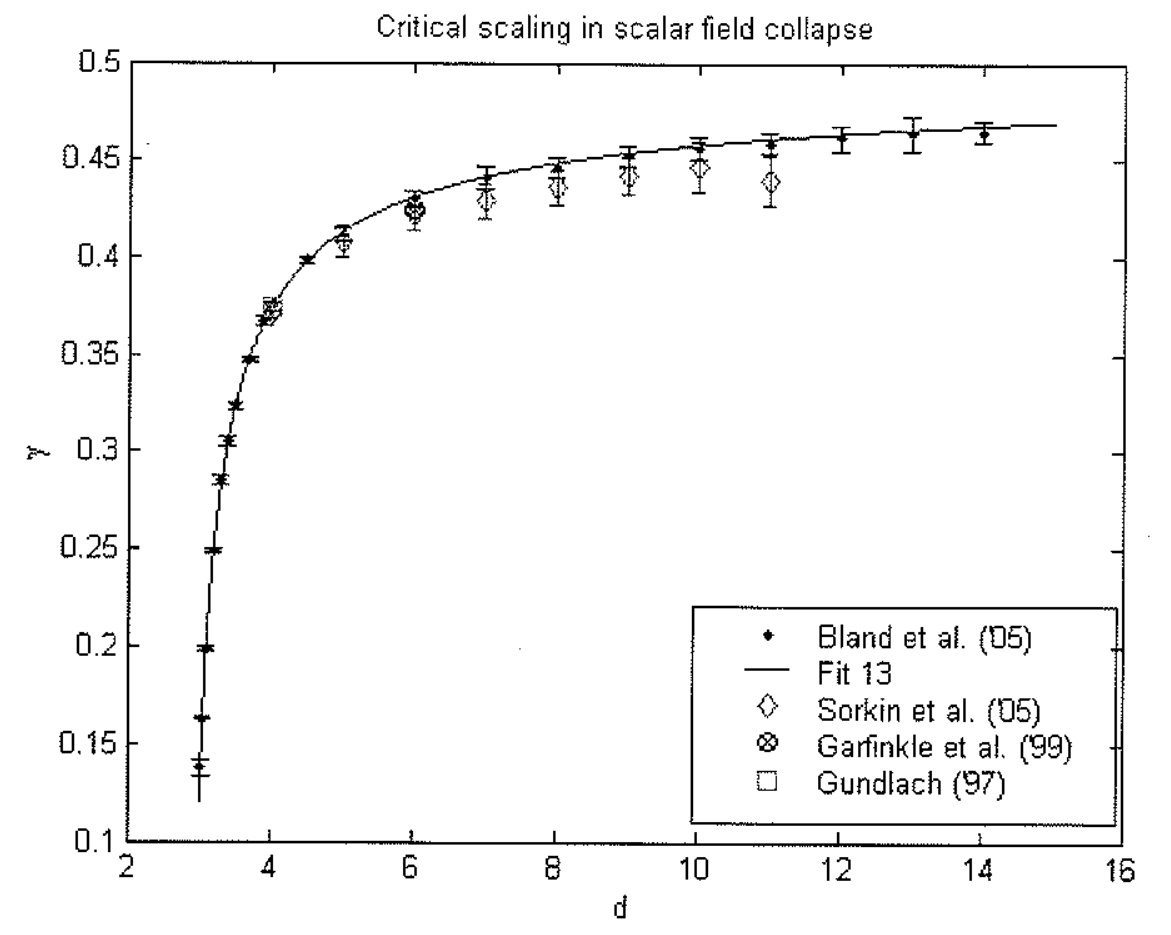}
\caption{Plot from \cite{blandthesis}, taking data from \cite{Garfinkle:1999zy,Sorkin:2005vz,Bland:2005vu,Gundlach:1996eg} showing the critical exponent $\gamma$ as a function of $D$ (appears as $d$ in figure).  The fit (labeled Fit 13) is to a function of the form $\gamma=\frac{1}{2}-\frac{m_3}{(D-m_2)^{m_1}}$.  }
\label{fig:bland}
\end{figure}

While this simple result is suggestive, it remains unclear whether $\gamma=1/2$ corresponds to the universal critical exponent expected at finite $D$. Intriguingly, studies of critical collapse in higher dimensions suggest that $\gamma\to1/2$ in the large $D$ limit.  We show in Fig. \ref{fig:bland} a plot from \cite{blandthesis} which demonstrates the trend towards $\gamma=1/2$. 

Some similarities can be found between our infinite $D$ solutions and the Roberts family of solutions \cite{Roberts:1989sk,Frolov:1998zt}. These are continuously self-similar solutions to the finite $D$ equations, including near the critical collapse threshold, which are known in closed form for $D=4$. The Roberts family resemble our oscillaton solutions in that they also contains solutions with or without horizons, connected continuously in parameter space.  Our solutions, however, do not exhibit self symmetry of any sort.

It was revealed in \cite{Frolov:1997uu,Frolov:1998tq} that the critical Roberts solution cannot correspond to the universal critical spacetime approached by fine tuning initial data. The reason is that perturbations of the Roberts solution contain more than one growing mode.  Additionally, it remains unclear whether the particular value of $\gamma=1/2$ given by the large $D$ oscillatons corresponds to the universal critical exponent expected at finite $D$. We do not attempt a perturbative analysis here, and leave that study for future work. 

\vspace{10pt}
\section{Large D scalars in $AdS$}
\label{intro}

We now develop the large $D$ limit of a scalar field in the asymptotically AdS case.

\vspace{10pt}
\subsection{Probe Scalar in $AdS$ Space}

Consider a probe massless scalar field in $AdS$ with the line element
\begin{equation}
\mathrm ds^2=\frac{L^2}{\cos^2 x}\left[-\mathrm dt^2+\mathrm dx^2+\sin^2 x\,\mathrm d\Omega_{D-2}\right]\;.
\end{equation}
With spherical symmetry, the solution to the scalar wave equation is
\begin{equation}\label{linearsol}
\phi(x,t)=\sum_{n=0}^{\infty}a_n\cos(\omega_n t+\phi_0)\cos^{D-1}(x) \, P_n^{(\frac{D-3}{2},\frac{D-1}{2})}(\cos 2x)\;,
\end{equation}
where $\omega_n=D-1+2n$, and $P_n^{(\alpha,\beta)}$ are Jacobi polynomials.  

We point out two important consequences of the large $D$ limit for those normal modes. First, the oscillations $\omega_n=D-1+2n$ become very rapid, which amounts to saying that low-lying modes oscillate very similarly to each other in time -- this is similar to the flat space case discussed previously. Second, the factor of $\cos^{D-1}x\equiv e^{-\rho}$, which exhibits the universal fall-off of any massless field in AdS, effectively divides AdS into two regions: $\rho\gg D$, where this factor decreases exponentially with increasing $D$, and $\rho\ll D$ where this factor decreases more slowly. For any $D$, the radius of the sphere $S^{D-2}$ dividing these regions remains $O(1)$. 
This suggests it is useful to think about the equations separately in both regions, which is our strategy below.

\vspace{10pt}
\subsection{Including Backreaction}

Let us now include backreaction with the metric ansatz
\begin{equation}
\mathrm ds^2=\frac{L^2}{\cos^2 x}\left[-(1-\delta)A\,\mathrm dt^2+\frac{\mathrm dx^2}{A}+\sin^2 x\,\mathrm d\Omega_{D-2}\right]\;.
\end{equation}
We will take the usual boundary condition $A=1$, $\delta=0$ at the conformal boundary $x=\pi/2$.  The equations of motion are given by 
\begin{subequations}
\begin{align}
\partial_tA&=-\frac{2}{D-2}\sin x\cos^D x\,A\,\partial_t\varphi\partial_x(\cos^{d-1}x\varphi),\\
\sin x\cos x\,\partial_xA&=(D-3+2\sin^2x)(1-A)-\sin^2x\cos^{2(D-1)}x S A\\
\partial_x\delta&=-2\sin x\cos^{2D-3}S(1-\delta)\\
0&=-\frac{1}{\sqrt{1-\delta}A}\partial_t\left(\frac{\partial_t\varphi}{\sqrt{1-\delta}A}\right)+\frac{\partial_x(\sin x\cos x\,\partial_x(\cos^{D-1}x\varphi))}{\sin x\cos^D x}+\nonumber\\
&\qquad\qquad+\frac{D-3+2\sin^2 x}{\sin x\cos^d xA}\partial_x(\cos^{D-1}x\varphi)\;,
\end{align}
\end{subequations}
where
\begin{equation}
S=\frac{1}{(D-1)(D-2)}\left[\left(\frac{\cos x\,\partial_t\varphi}{\sqrt{1-\delta}A}\right)^2+\left(\frac{\partial_x(\cos^{D-1}x\,\varphi)}{\cos^{D-2}x}\right)^2\right]\;.
\end{equation}

Let us assume that the separation of regions discussed for the probe scalar is preserved in the large $D$ limit, even when the backreaction is included.  That is, we take $\phi=\cos^{(D-1)}x\,\varphi$, where $\varphi$ does not scale with $D$. We also change the radial coordinate $\cos^{D-1}x\equiv e^{-\rho}$. 
In the outer region $\rho\gg D$, the rapid fall-off of $e^{-\rho}$ decouples the scalar field from the metric, and the solution is essentially given by \eqref{linearsol}.  In the inner region $\rho\ll D$, backreaction must be taken into account, and the equations are nonlinear. Boundary conditions for the inner equations should be supplied from matching to the outer solution.  

Let us discuss the two regions more precisely. In the outer region, where $\rho\gg D$, the exponential fall-off decouples the metric and reduces the equations to the linear scalar field equation
\begin{equation}\label{lineq}
-\partial_t^2\varphi-\partial_x^2 \varphi-2(1-(D-1)\cos 2x)\csc2x\,\partial_x\varphi-(D-1)^2\varphi=0\;,
\end{equation}
while the metric remains $A=1$, $\delta=0$.  

As we have mentioned, the solution to the linear equation \eqref{lineq} is given by Jacobi polynomials \eqref{linearsol}, for every $D$. We can also solve the linear equations \eqref{lineq} order by order in a power series expansion in $D^{-1}$.  The equations to lowest two orders  are solved by 
\begin{equation}\label{outermodes}
\varphi=\sum_n a_n \cos{[(D-1+2n)t+\phi_n]}\cos^{n} 2x+O(D^{-1})\;,
\end{equation}

We now turn to the inner region, where it is convenient to use the $\rho$ coordinate, where $\cos^{D-1}x\equiv e^{-\rho}$ and $\rho\ll D$. Let us also transform to a new time coordinate given by $\tau=(D-1)t.$  Let us also expand the functions in a power series in $D^{-1}$
\begin{equation}
A=1-\frac{1}{D}A_0+O(D^{-2})\;,\qquad \delta=\delta_0+O(D^{-1})\;,\qquad \varphi=\varphi_0+O(D^{-1})\;,
\end{equation}
where the lowest order term in $A$ is determined by the equations of motion. 

At leading order in the large $D$ expansion, we obtain the nonlinear equations in the inner region
\begin{subequations}\label{innereq}
\begin{align}
A_0&=\frac{2\rho e^{-2\rho}(\partial_\tau\varphi_0)}{1-\delta_0}\;,\\
\partial_\rho \delta_0&=-2e^{-2\rho}(\partial_\tau\varphi_0)^2\;,\\
\partial_\rho\varphi_0&=\frac{1}{\sqrt{1-\delta_0}}\partial_\tau\left(\frac{\partial_\tau\varphi}{\sqrt{1-\delta_0}}\right)+\varphi_0\;.
\end{align}
\end{subequations}
We note that the scalar equation is parabolic in character, but with two derivatives in $\tau$, and one derivative in $\rho$.  Thus, it is most naturally solved by declaring some `initial data' at particular value of $\rho$, and then integrating in $\rho$. This is similar to the method of solution discussed above, in the asymptotically flat case.

\vspace{10pt}
\subsection{Matching Solutions}
A full solution must solve both the inner and outer equations and match in an appropriate way. We observe that at large $\rho$, the inner equations reduce to the decoupled equations
\begin{equation}\label{farphi}
\partial_\rho\varphi_0=\partial_\tau^2\varphi_0+\varphi_0\;,
\end{equation}
and
\begin{equation}\label{fardelta}
\delta_0=0\;,\qquad A_0=0\;,
\end{equation}
where we have used our boundary condition for $\delta$. 
The scalar equation \eqref{farphi} describes the scalar field in the asymptotia of the inner region. It is equivalent to the  scalar equation linearised about empty $AdS$. The general solution of \eqref{farphi} is given by
\begin{equation}
\varphi_0=\int\mathrm d\omega \, \alpha(\omega) \, e^{i\omega\tau-(\omega^2-1)\rho}\;,
\end{equation}
for some frequency distribution $\alpha(\omega)$ that obeys the reality condition for $\varphi_0$. For the purpose of matching with solutions in the outer region, we choose $\alpha(\omega)$ to be a sum of delta functions, i.e. a discrete sum of modes, then
\begin{equation}\label{farmodesol}
\varphi_0=\sum_n \alpha^{\pm}_ne^{\pm i n\tau-(n^2-1)\rho}\;.
\end{equation}
Note that the mode $n=0$ diverges and, as we shall see, cannot be matched to a regular solution to the outer equations. 

Now let us consider the outer solution, which  is given as a sum of modes by \eqref{linearsol}. To match to the large $\rho$ limit of a solution of the inner scalar equations, whose general form is \eqref{farmodesol}, we must translate between different conventions in our treatment of the inner and outer regions.  First we change  to the $\rho$ and $\tau$ coordinates, which, recall are defined by $\cos^{D-1}x\equiv e^{-\rho}$ and $\tau=(D-1)t$. Furthermore, the outer solution \eqref{linearsol} is exact in $D$, thus to match at a given order in the large $D$ expansion we need to expand it in a power series in $D^{-1}$ which is given by \eqref{outermodes}.  Then after the coordinate transformation, an expansion in $D^{-1}$ gives
\begin{equation}\label{outermatch}
\varphi=\sum_{\pm}e^{\pm i\tau}\left[E^{\pm}_0+O(D^{-1})\right]
\end{equation}
where for any integer $k$ we define
\begin{equation}\label{Edef}
E^{\pm}_k=\sum_n a^{\pm}_n n^{k}\;.
\end{equation}
The $E_k$ for $k>0$ appear in the expansion \eqref{outermatch} at higher orders in $D^{-1}$.

The scalar field \eqref{outermatch} is then a solution to the large $\rho$ limit of the inner equations \eqref{farphi} which matches a general multi-mode solution to the outer equations \eqref{outermodes}, to leading order in the large $D$ expansion. Thus, the configuration \eqref{outermatch} can be supplied as boundary conditions for the inner equations at large $\rho$.  

Importantly, although the outer multi-mode solution is parametrised by the amplitudes $a^{\pm}_n$ for each mode, the inner equations only use the combinations of these amplitudes that appear in the $E^{\pm}_k$. In particular,  the leading order inner solution only uses $E_0$, the sum of the amplitudes.  Therefore, at any fixed order in the large $D $ expansion there are many outer solutions that have the same inner solution.  

Conversely, there are also many inner solutions that can be matched to the same outer solution.  Recall that the large $\rho$ solution of the inner equations, written as a sum of modes with ampltiudes $\alpha^\pm_n$ is given by \eqref{farmodesol}.  Matching to the form \eqref{outermatch} at leading order only determines $\alpha^\pm_n$ for $n=0$ and $n=1$. The $n=0$ mode must vanish, and the $n=1$ mode must match $E_0^\pm$. All $n>1$ fall off exponentially with $\rho$ and thus can be consistently matched to \eqref{farmodesol}.

This matching introduces an importance difference between the flat space case and the AdS case.  In the flat space case, we can integrate any `initial data' at $r=0$ outwards to large $r$, where the scalar field must decay. In the AdS case, the same sort of calculation for the inner equations would generically lead to a nonzero $n=0$ mode at large $\rho$, and thus cannot be matched to a solution to the outer equations. This matching therefore constrains the configurations of the scalar field at the origin.

\vspace{10pt}
\subsection{Higher Orders}

For completeness, let us also describe the matching at higher orders in the large $D$ expansion.  First, let us obtain a higher-order solution to the linear equation \eqref{lineq}:
\begin{align}\label{outermodes2}
\varphi=\sum_n& a_n \cos{[(D-1+2n)t+\phi_n]}\bigg\{\cos^{n} 2x-\nonumber\\
&\qquad-\frac{1}{D}\left(n \cos^{n-1}2x+\frac{1}{2}n(n-1)\cos^{n-2}2x\right)\nonumber\\
&\qquad+\frac{1}{D^2}\bigg(2n(n-1)\cos^{n-1}2x+n(n-1)^2\cos^{n-2}2x+\frac{1}{2}n(n-1)(n-2)\cos^{n-3}2x+\nonumber\\
&\qquad\qquad+\frac{1}{8}n(n-1)(n-2)(n-3)\cos^{n-4}2x\bigg)+O(D^{-3})\bigg\}\;.
\end{align}
Now we perform the coordinate transformation to $\rho$ and $\tau$ defined by $\cos^{D-1}x\equiv e^{-\rho}$ and
\begin{equation}
\tau=\left(D-1+\sum_{k=0}^{\infty}\frac{\lambda_k}{D^k}\right)t\;,
\end{equation}
where the $\lambda$'s are constants of our choosing. If we take the phases to be aligned $\phi_n=0$, then this coordinate transformation and another $D^{-1}$ expansion yields
\begin{align}\label{outerexpand}
\varphi&=E_0\cos\tau+\frac{1}{D}\left[-\left(\frac{1}{2}(E_1+E_2)-4E_1\rho\right)\cos\tau+\tau E_0\left(\lambda_0-2\frac{E_1}{E_0}\right)\sin\tau\right]+\nonumber\\
&\qquad+\frac{1}{D^2}\bigg\{-\bigg[\frac{1}{8}(6E_1+E_2-6E_3-E_4)+2(2E_1+E_2-E_3)\rho+4(E_1-2E_2)\rho^2+\nonumber\\
&\qquad\qquad\qquad\qquad+\frac{1}{2}\left(4E_2+\lambda_1E_0\left(\lambda_0-4\frac{E_1}{E_0}\right)\right)\tau^2\bigg]\cos\tau+\nonumber\\
&\qquad\qquad\qquad+\tau\bigg[(1-\lambda_0)E_0\left(\lambda_0-2\frac{E_1}{E_0}\right)+E_0\left(\lambda_1+\frac{E_2+E_3}{E_0}-\frac{\lambda_0(E_1+E_2)}{2E_0}\right)-\nonumber\\
&\qquad\qquad\qquad\qquad-4E_1\left(\lambda_0-2\frac{E_2}{E_1}\right)\rho\bigg]\sin\tau\bigg\}\;,
\end{align}
where the $E$'s have been defined analogously to \eqref{Edef}.  The large $\rho$ limit of the inner equations must agree with \eqref{outerexpand} in order to form a full solution. 

We see that at higher orders, there are terms that are polynomial in $\tau$. Some of these terms can be removed by a choice for $\lambda_k$, but not all of them, unless some extra constraints are placed on the $E$'s.  This expansion would imply that there is a breakdown of the large $D$ expansion when $\tau\sim O(D)$.  But we know that these outer solutions can be formed from the solution \eqref{linearsol} or the expansion \eqref{outermodes2} which do not contain any divergences.  So in this case, we know that there exists a resummation of the perturbation series to a well-behaved solution.  It is less clear whether or not the inner solutions, which inevitably need to be matched to the series \eqref{outerexpand}, can similarly be resummed.  

We now consider a particular restriction on $E_k$ that removes the polynomial terms in $\tau$ that appear at this order. Let us assuume $E_0\neq0$ and then take $\lambda_0=2E_1/E_0$, and $\lambda_1=[E_0(E_1+E_2)-E_0(E_2+E_3)]/E_0^2$ to remove some of these terms.  Then the remaining terms that are polynomial in $\tau$ are all proportional to 
\begin{equation}
E_0E_2-E_1^2=\frac{1}{2}\sum a_na_m(n-m)^2\;.
\end{equation}
Note that this is zero for single-mode data, and (generically) nonzero for multi-mode data. We therefore have a connection to the AdS instability, where single-mode dominated data is expected to be stable, and multi-mode data is expected to form black holes.

In finite $D$, if one performs perturbation theory about $AdS$ with a scalar field of amplitude $\epsilon$, there is a secular term that appears which leads to breakdown of perturbation theory at $t\sim1/\epsilon^2$.  This secular term appears for any multi-mode data, and is absent for single-mode data. It seems the large $D$ expansion in AdS is exhibiting analogous behaviour. 

\vspace{10pt}
\subsection{Oscillatons}
There is  an analytic solution to \eqref{innereq} given by
\begin{equation}
\varphi_0=e^{\rho}\tanh^{-1}(E_0\,e^{-\rho}\sin(\tau-\tau_0))\;,\qquad \delta_0=\frac{[E_0\,e^{-\rho}\cos(\tau-\tau_0)]^2}{1-[E_0\,e^{-\rho}\sin(\tau-\tau_0)]^2}\;,
\end{equation}
for constants $E_0$ and $\tau_0$. The constant $E_0$ matches the value of $E_0$ as defined above for the outer solution.  The function $A_0$ is given by
\begin{equation}
A_0=\frac{2E_0^2\,\rho\,e^{-2\rho}\cos^2(\tau-\tau_0)}{(1-E_0^2\,e^{-2\rho})(1-E_0^2\,e^{-2\rho}\sin^2(\tau-\tau_0))}\;.
\end{equation}

Matching this inner solution to an outer solution consisting of a single mode, the global solution oscillates with a single period, and therefore represents an oscillaton at large $D$. Like finite $D$ oscillatons, there is an upper bound to their amplitude, which here is signald by the divergence of the inner solution, occurring when $E_0\geq1$. As we discussed in the asymptotically flat context, we conjecture that this divergence is indicative of horizon formation at finite $D$.

Note that this inner solution can just as well be matched with outer solutions consisting of multiple modes, so long as they generate the same $E_0$. This suggests that at finite but large $D$, multi-mode initial data with $E_0<1$ are especially long-lived compared with those with $E_0>1$. This introduces a some tension with results at finite $D$. Numerical evidence at finite $D$ suggests that initial data sufficiently far from a normal mode (i.e. not single-mode dominated) will eventually collapse to form a black hole.  Yet, the solutions at large $D$ we have found includes, for example, equal-amplitude two-mode initial data which is apparently long lived at large $D$, but is expected to collapse at finite $D$.  

We suspect this conflict arises as a consequence of an order of limits, i.e. the large $D$ limit does not necessarily commute with the long time limit. This can be seen more explicitly by the terms that appear at higher order \eqref{outerexpand} that are polynomial in $\tau$.  Indeed, the fact that single-mode data is free from these terms (after a suitable choice of time coordinate), and that multi-mode data generically has them is consistent with expectations from studies of the AdS instability.  Nevertheless, the fact that these terms only show up at higher order suggests that collapsing data at finite $D$ with a fixed $E_0<1$ will take longer and longer to collapse as $D$ is increases.

\vspace{10pt}
\subsection{Boson Stars}
We now wish to compare the infinite-$D$ solutions to those at finite $D$.  However, oscillatons require the numerical solution to a nonlinear PDE, which becomes difficult to compute when $D$ is large.  Instead, we opt to make a comparison to boson stars, which use a complex instead of a real scalar.  The complex scalar field is chosen to be of the form $\varphi=e^{i\omega t}\psi$, where $\psi$ is a real function of the the radial coordinate. The time dependence in the phase cancels out in the metric, allowing the solution to be solved as an ODE. At leading order in the large $D$ limit, the equations of motion become
\begin{subequations}\label{bstarinnereq}
\begin{align}
A_0&=\frac{4\rho e^{-2\rho}\psi_0}{1-\delta_0}\;,\\
\partial_\rho \delta_0&=-4e^{-2\rho}\psi_0^2\;,\\
\partial_\rho\psi_0&=-\frac{\delta_0}{1-\delta_0}\psi_0\;.
\end{align}
\end{subequations}
Note that $\omega$ dependence drops out of the equations at leading order since $\omega\sim (D-1)+2k$. Though the above equations are just ODEs, we unfortunately do not have an analytic solution to the above equations. However, they can be straightforwardly solved numerically.  The boundary conditions at large $\rho$ given by $\delta_0=0$, $A_0=0$, and $\psi_0=E_0$ for some constant $E_0$ can be obtained by matching to the outer solution. We solve these equations numerically in the coordinate $z=\sqrt{1-e^{-\rho}}$, which has a finite range $z\in[0,1]$.

To compare with boson stars at finite $D$, we must choose a family.  Like oscilltons, when scalar field of the boson stars are perturbatively small, the frequency is of the form $\omega=D-1+2k$ for some non-negative integer $k$.  Each choice of $k$ produces a different one-parameter family of boson stars.  Since we do not assume $k$ scales with $D$ in the large $D$ limit, it suffices to take the $k=0$ family. 

\begin{figure}[h]
\centering
\includegraphics[width=.5\textwidth]{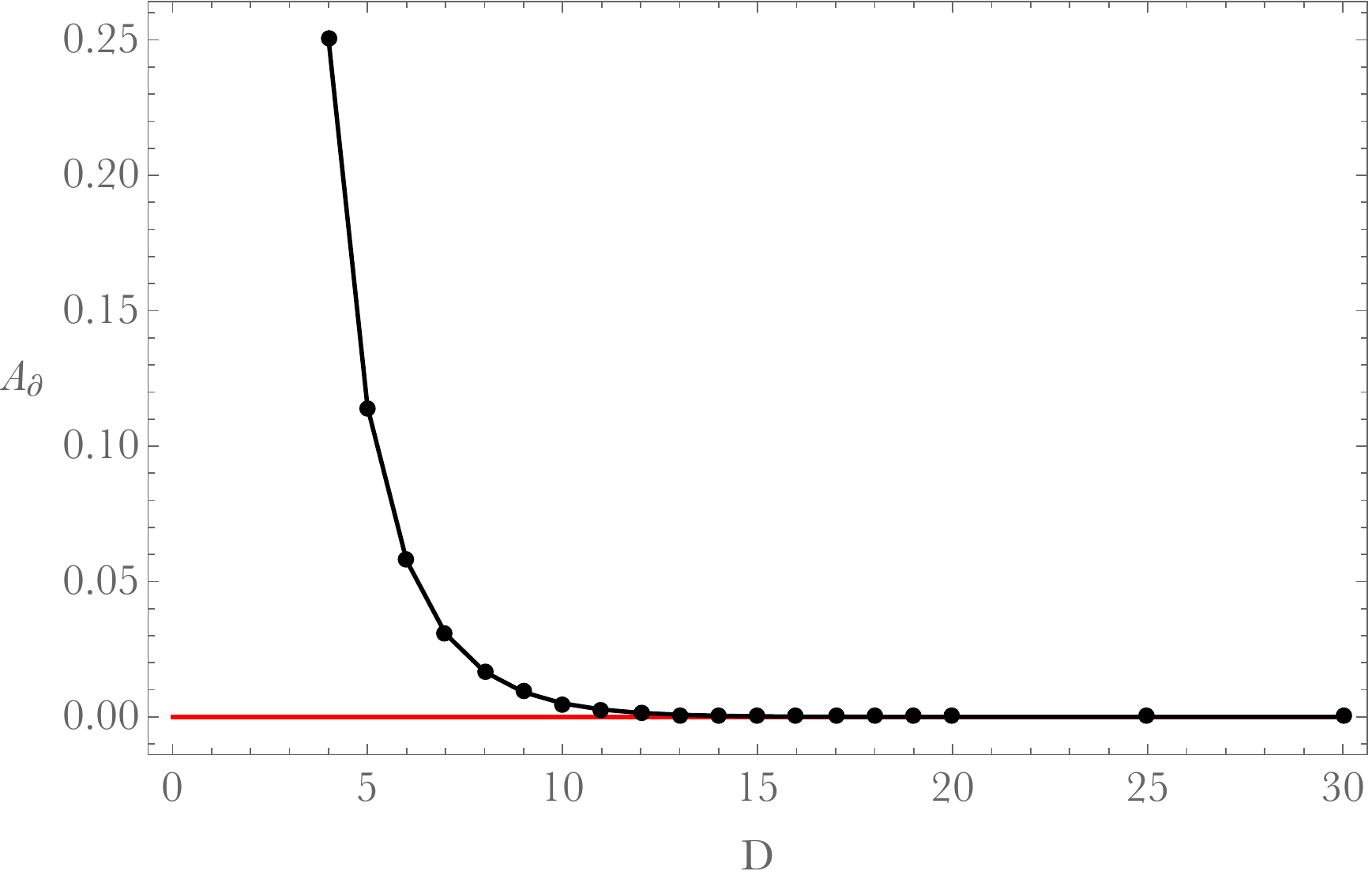}
\includegraphics[width=.5\textwidth]{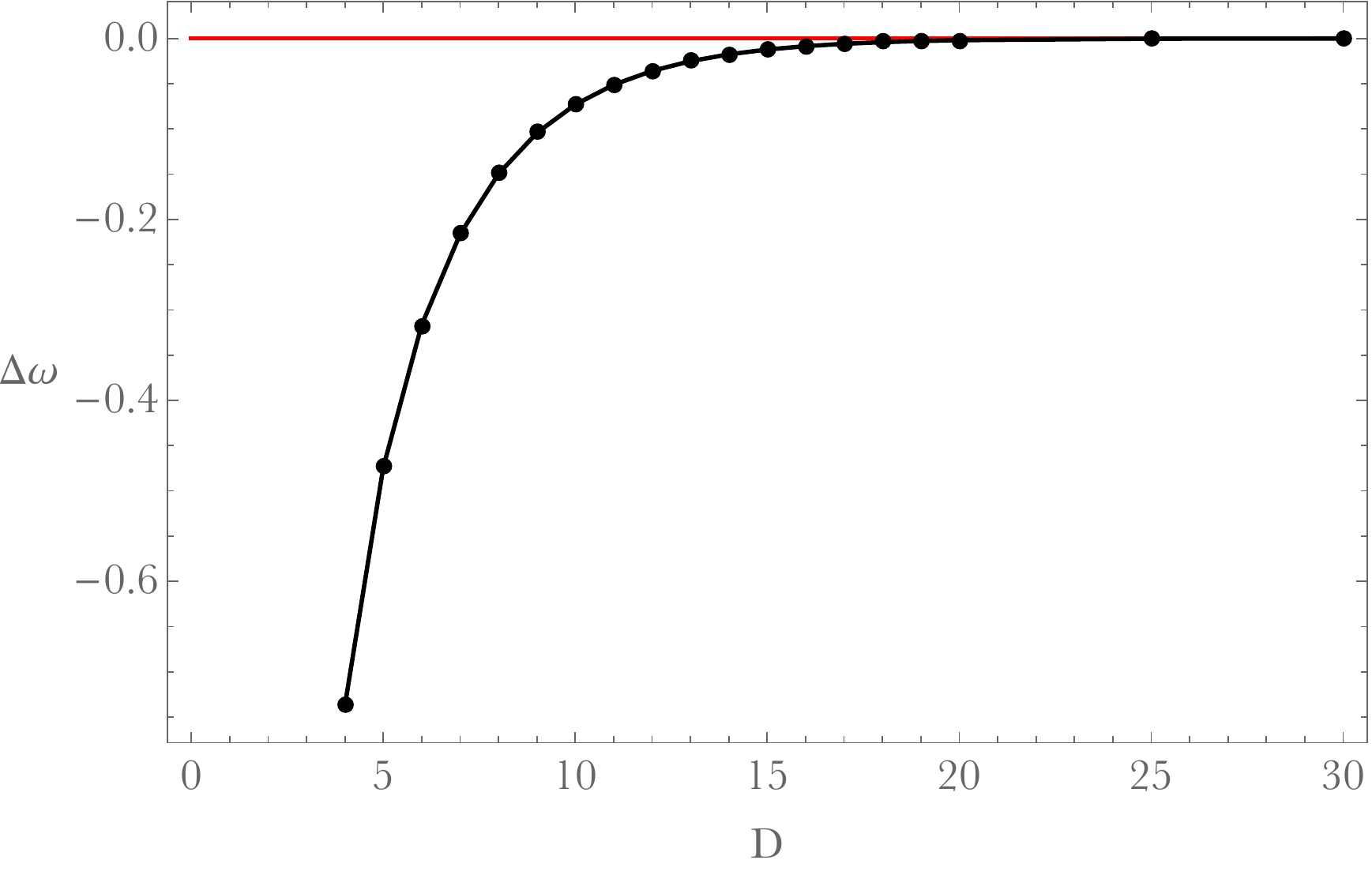}
\includegraphics[width=.5\textwidth]{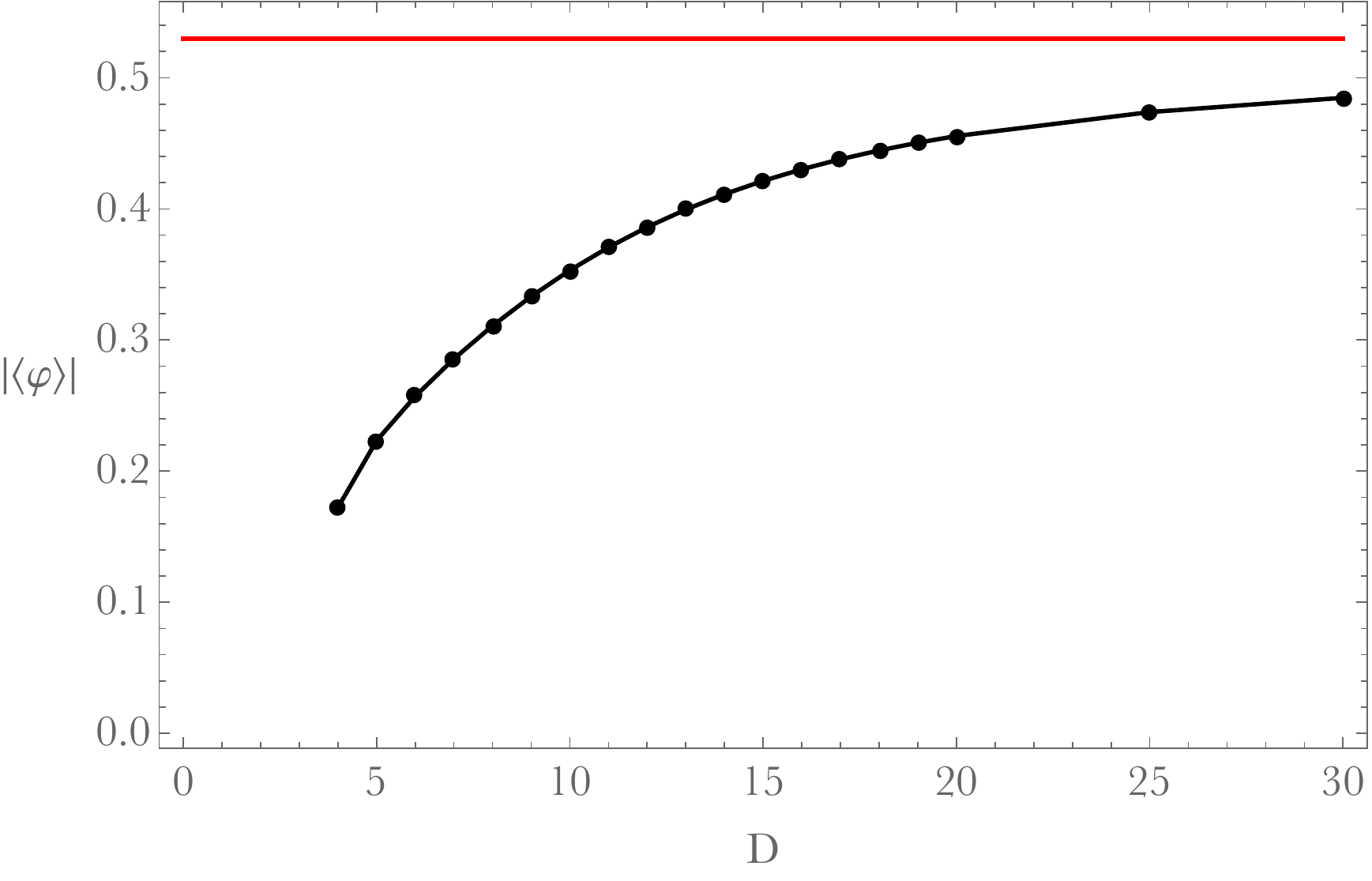}
\caption{Comparison of stars at finite $D$ (black dots) to those from the leading order large $D$ limit (red line). All solutions have $\psi(0)=400$. Here, $A_\partial\equiv A(x=p/2)$ is proportional to the energy, $\Delta\omega\equiv \omega-(D-1)$ is the difference in frequency to from that of the perturbative normal mode, and $|\langle\varphi\rangle|\equiv\psi(x=\pi/2)$ is the value of the scalar field at the boundary (related via AdS/CFT to the expectation value of a scalar operator).}
\label{fig:bosonstars}
\end{figure}

Parametrise the boson stars by the value of the scalar field at the origin $\psi(0)$.  We choose $\psi(0)=400$ in order to compare highly nonlinear boson stars.  We compare three quantities: $A_\partial\equiv A(x=p/2)$ (which is proportional to the energy), $\Delta\omega\equiv \omega-(D-1)$, and $|\langle\varphi\rangle|\equiv\psi(x=\pi/2)$. The first two these, $A_\partial$ and $\Delta\omega$ vanishes at all orders in the large $D$ expansion.  The results as a function of $D$ are shown in Fig. \ref{fig:bosonstars}.

We see from Fig. \ref{fig:bosonstars} that $A_\partial$ and $\Delta\omega$ rapidly approach zero (the large $D$ value) as $D$ increases, while $|\langle\varphi\rangle|$ approaches the large $D$ value more slowly.  This is unsurprising due to the fact that $|\langle\varphi\rangle|$ receives corrections at higher orders while the other quantities do not.  

Since linear perturbations of boson stars also involve only ODEs, let us attempt a stability analysis.  We take $\psi=\psi_0+\epsilon[\cos(\lambda\tau)\delta\psi_r+i\,\sin(\lambda\tau)\delta\psi_i]$, as well as $\delta=\delta_0+\epsilon\,\cos(\lambda\tau)\delta\delta$ and expand the leading order large $D$ equations to linear order in $\epsilon$, taking $\delta\psi_r$, $\delta\psi_i$, and $\delta\delta$ to be functions of $z=\sqrt{1-e^{-\rho}}$. The resulting linear equations are

\begin{subequations}
\begin{align}
(1-z^2)(1-\delta_0)\partial_z\delta\psi_r&=-2z\left[\frac{(1-z^2)^2\varphi_0}{1-\delta_0}\delta\delta+\delta_0\delta\psi_r+2\lambda\delta\psi_i+\lambda^2\delta\psi_r\right]\\
(1-z^2)(1-\delta_0)\partial_z\delta\psi_i&=-2z\left[\delta_0\delta\psi_i+2\lambda\left(\delta\psi_r+\frac{(1-z^2)^2\varphi_0}{4(1-\delta_0)}\delta\delta\right)+\lambda^2\delta\psi_i\right]\\
(1-z^2)\partial_z\delta\delta&=4z\left[\delta\delta-4\varphi_0(\delta\psi_r+\lambda\delta\psi_i)\right]\;.
\end{align}
\end{subequations}
It turns out that a series around $z=1$ yields an algebraic equation for $\lambda$ which depends on how quickly $\psi_r$ and $\psi_i$ vanish. For example, choosing $\psi_r$ and $\psi_i$ to vanish linearly around $z=1$ implies $1 - 6 \lambda^2 + \lambda^4=0$, which gives the lowest non-trivial frequency we have found: $\lambda=\sqrt2-1$.  These frequencies, and others like it, are independent of the background solution $\phi_0$, $\delta_0$.  Numerically solving the linear equations do not reveal any additional frequencies that depend on the background solution.  This suggests that the boson stars are linearly stable at large $D$. 

\vspace{20pt}
\section{Conclusions and Outlook}
We have developed the large $D$ limit for spherically symmetric scalar fields in both flat space and in AdS, arriving at an effective set of large $D$ equations.  These equations are parabolic in character, and may be solved by starting with data at the origin (for all time), and then integrating outwards to large radii.  In the AdS case, the solution must additionally be matched to an outer region, which places restrictions on the set of data that can be supplied at the origin. This matching at higher orders also requires terms that are polynomial in time $\tau$ which may possibly cause the breakdown of the perturbation series when $\tau\sim O(D)$, which can be analogously compared to the $t\sim 1/\epsilon^2$ timescale where perturbation theory in AdS (at any finite $D$) breaks down.

In both flat space and AdS, we have found analytic solutions to the leading order equations that represent scalar fields oscillating at a single frequency. These solutions form a one-parameter family, and are regular up until this parameter reaches a critical value, at which the scalar field contains a divergence. While black hole collapse does not occur in the large $D$ limit, interpreting this divergences as finite $D$ horizon formation leads to a critical exponent of $\gamma=1/2$.   While we have not demonstrated whether or not this exponent is related to the universal exponent for critical collapse at finite $D$, the value of $\gamma=1/2$ appears consistent with extrapolations of finite $D$ results.   If, indeed there is a connection between the solutions we have found and critical collapse, that would imply that there is a connection at finite $D$  between the singular limit of oscillatory solutions and the critical solution of gravitational collapse.  

Though asymptotically flat oscillatons do not exist at finite $D$ for massless fields, we have found them in the large $D$ limit. This may be a consequence of the effective equations becoming massless in this limit. At finite $D$, there may be long-lived configurations of massless scalars that resemble oscillatons. 

We have also constructed AdS boson stars at large $D$ numerically. A linear stability analysis yields a number of frequencies that do not depend on the background solution, and suggests that these solutions are linearly stable. 

In this work, as in many other studies of the large $D$ limit, we have made heavy use of spherical symmetry. It should be possible to construct effective large $D$ field equations that break several symmetries of the sphere. The effective large $D$ equations we have obtained largely relies on the rapid falloffs of the scalar field at large $D$, and these falloffs should be present even when spherical symmetry is broken.  These falloffs are present for many other matter fields as well, and even for gravitons, so we expect these methods to be applicable to a wide variety of matter content. 

In taking this large $D$ limit, we have required the time coordinate to be rescaled with $D$. We have also chosen a particular gauge for the metric.  The metric we have chosen cannot have the continuous or discrete self-similarity exhibited by the finite $D$ critical solution at the threshold of collapse.  While the critical solution may have a non-self-similar description in this large $D$ limit, it is unclear what form this particular solution would take. Perhaps an alternative gauge or scaling would allow for the critical solution, and perhaps other solutions as well, to be constructed at large $D$.  

It would be interesting to see if this horizonless large $D$ limit can be joined with the well-developed large $D$ limit of black holes. The resulting theory may describe situations where the interactions between the matter and the black hole are important.   We note that the time scaling we have taken for the asymptotically flat case $t\sim1/\sqrt{D}$ has not yet appeared in any large $D$ study of black holes\footnote{We thank Roberto Emparan for pointing this out.}, and may be an important time scale in the dynamics of critical collpase. 

\vspace{20pt}
\section*{Acknowledgements}
It is a pleasure to thank Matthew Choptuik, Roberto Emparan, and Shiraz Minwalla for helpful comments. We are supported by a discovery grant from NSERC.

\vspace{20pt}
\bibliographystyle{JHEP}
\bibliography{bibl}

\end{document}